\documentclass[%
twocolumn,
showpacs,
 amsmath,amssymb,
 aps,prc
]{revtex4-1}

\usepackage{graphicx}
\usepackage{dcolumn}
\usepackage{bm}
\usepackage{color}

\begin{document}

\title{Coherent pion production in neutrino-nucleus scattering}

\author{Kapil Saraswat$^1$}
\author{Prashant Shukla$^{2,3}$}
\email{pshuklabarc@gmail.com}

\author{Vineet Kumar$^2$}
\author{Venktesh Singh$^1$}

\affiliation{$^1$Department of Physics, Institute of Science, Banaras Hindu University, Varanasi 221005, India.}
\affiliation{$^2$Nuclear Physics Division, Bhabha Atomic Research Centre, Mumbai  400085, India.}
\affiliation{$^3$Homi Bhabha National Institute, Anushakti Nagar, Mumbai 400094, India.}

\date{\today}

\begin{abstract}
  In this article, we study the coherent pion production in the neutrino-nucleus interaction
in the resonance region using the formalism based on the partially conserved axial current (PCAC) 
theorem which relates the neutrino-nucleus cross section to the pion-nucleus elastic cross section. 
 The pion nucleus elastic cross section is calculated using the Glauber model in terms of
pion-nucleon cross sections obtained by parameterizing the experimental data. 
  We calculate the differential and integrated cross sections for charged current coherent
pion production in neutrino carbon scattering. The results of integrated cross-section 
calculations are compared with the measured data.
  Predictions for the differential and integrated cross sections for coherent pion 
productions in neutrino-iron scattering using above formalism are also made.
\end{abstract}




\maketitle


\section{Introduction}

  The neutrinos generated in the upper atmosphere are the best tools for studying the 
phenomena of neutrino oscillations considered by many experiments planned worldwide 
\cite{Ashie:2005ik, Takeuchi:2011aa, Aliu:2004sq, Ahn:2002up, Ahn:2006zza, Ahmed:2015jtv}.
For a review of neutrino oscillation experiments see Ref.~\cite{Messier:2006yg}.
 The flux of atmospheric neutrinos rapidly falls with energy \cite{Honda:2011nf}.
 Typically, the detectors measure recoil muons which are produced by charged current 
interaction of neutrinos inside the detector medium, e.g., carbon. In work at the 
India-Based Neutrino Observatory (INO), the medium is iron \cite{Ahmed:2015jtv}.
  The intermediate-energy neutrinos (between energies of 1 and 3 GeV) produce the 
bulk of the signal. 
  The neutrino interaction with matter in the intermediate-energy range gets 
contribution from many processes which include quasi elastic scattering,
interaction via resonance pion production, and deep inelastic scattering \cite{Formaggio:2013kya}. 
One of the most important processes in the resonance production region is 
coherent pion production. In this process, the nucleus interacts as a
whole with the neutrino and remains in the same quantum state as it was initially 
before the interaction. 
 It happens when the four-momentum transfer $|t|$ to the nucleus remains small. 
The characteristic signature of coherent scattering is a sharp peak in the low $|t|$ region. 
The cross section for this process is proportional to the square of the mass number 
of the nucleus and has logarithmic dependence on neutrino energy \cite{Paschos:2005km}.
  Coherent pion production plays an important role in the analysis of data of neutrino 
oscillation experiments. The neutral current $\pi^{0}$ production is one of the largest 
background sources in muon-neutrino experiments 
\cite{AguilarArevalo:2007it, AguilarArevalo:2008xs} 
because one of the photons from $\pi^{0}$ can mimic electron events. 
The charged current coherent pion production was one of the most rigorously 
studied background processes in K2K experiment \cite{Hasegawa:2005td}. 
  Charged current coherent pion production has been observed experimentally at higher 
energies and for several nuclei 
\cite{Grabosch:1985mt, Marage:1986cy, Allport:1988cq, Aderholz:1988cs, Willocq:1992fv, Vilain:1993sf}. 

  The most common theoretical  approach for describing coherent pion 
production is based on Adler's Partially Conserved Axial Current (PCAC) 
theorem which relates the neutrino induced coherent pion production to 
the pion-nucleus elastic scattering 
\cite{Rein:1982pf, Belkov:1986hn, Kopeliovich:2004px, Gershtein:1980vd, Komachenko:1983jv, Paschos:2005km, Berger:2008xs, Paschos:2009ag}. 
 The PCAC model has been successful in 
describing coherent pion production at high energy \cite{Rein:1982pf, Belkov:1986hn}.
Work with the same assumption has been used at low energy in Ref. \cite{Paschos:2005km}. 
There are several microscopic models as well for coherent scattering,
e.g., in Refs. \cite{Kim:1996az, Kelkar:1996iv, Singh:2006bm, AlvarezRuso:2007tt, AlvarezRuso:2007it, Amaro:2008hd}.

  In this work, we calculate the differential and integrated cross sections for charged current 
neutrino nucleus coherent scattering using the PCAC-based model developed by 
Kopeliovich and Marage \cite{Kopeliovich:1992ym}.
 The pion nucleus elastic cross section is calculated using the Glauber model in terms of
pion-nucleon cross sections obtained by parametrizing the experimental data. 
  The differential and integrated cross sections for charged current 
neutrino-carbon coherent scattering are calculated and compared with the measured data.
 Predictions for the differential and integrated cross sections for neutrino-iron 
coherent scattering using the formalism above are also given.

\section{PCAC Based Model}
The scattering process of charged current coherent pion production is 
given as 
\begin{equation}
\nu_{\mu}(k) +A(p) \rightarrow \mu^{-}(k^{'})+\pi^{+}(p_{\pi})+A(p^{'}).
\label{cccoherentpionproductionequation}
\end{equation}

\begin{figure}
\includegraphics[width=0.70\linewidth]{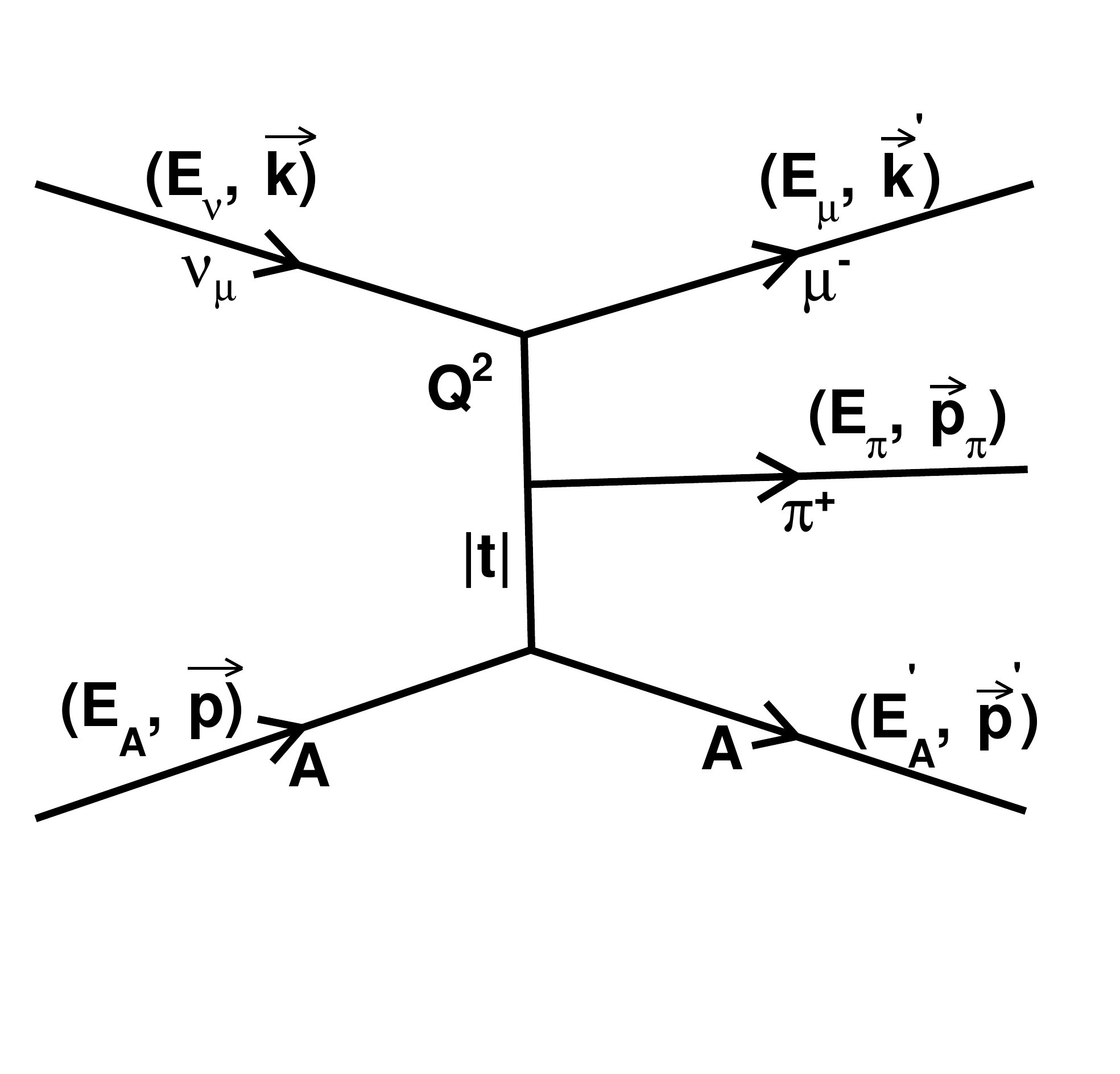}
\caption{Charged current coherent pion production.}
\label{fig1coherent}
\end{figure}

The schematic diagram of coherent pion production is shown in
Fig.~\ref{fig1coherent}. Here $k$ and $k^{'}$ are the 4-momenta of the incoming 
neutrino and outgoing lepton, respectively, and $p_{\pi}$ is that of the produced pion. 
If $E_{\nu}$ and $E_{\mu}$ are the energies of incident neutrino and outgoing lepton,
respectively, then $\nu~(=E_{\nu}-E_{\mu})$ is the energy and $q~(=k-k^{'})$ is the 
4-momentum transfer between the incoming neutrino and outgoing lepton. The momentum
transfer $Q^{2}$ is calculated as: $Q^{2}=-q^{2}={\bf q}^{2}-\nu^{2}$ .
 The features of the coherent pion production are characterized 
by the variable $|t|$ which is the squared momentum 
transfer to the nucleus from the neutrino-pion system
\begin{equation}
|t|=\Big|\Big(q-p_{\pi}\Big)^{2}\Big| = \Big|\Big(k-k^{'}-p_{\pi}\Big)^{2}
\Big|~.
\end{equation}
In coherent scattering $|t|$ remains small. 

The differential cross section for the charged current 
coherent pion production scattering process
\cite{Kopeliovich:1992ym, Berger:2008xs} is 
\begin{eqnarray}
\frac{d\sigma^{CC}}{dQ^{2} d\nu dt} &=& \frac{G^{2}_{F} \cos^{2}\theta_{C} f^{2}_{\pi}}
{2 \pi^{2}} \frac{u v}{|{\bf q}|}~
\Bigg[\Big(G_{A}-\frac{1}{2} \frac{Q^{2}_{m}}
{(Q^{2}+m^{2}_{\pi})}\Big)^{2}  \nonumber  \\
&~& + \frac{\nu}{4 E_{\nu}} (Q^{2}-Q^{2}_{m}) \frac{Q^{2}_{m}} 
{(Q^{2}+m^{2}_{\pi})^{2}}\Bigg]   
 \times \frac{d\sigma(\pi^{+}A \rightarrow \pi^{+}A)}{dt}~.
\label{marage1993}
\end{eqnarray}
Here $G_{F}$ (=1.16639 $\times 10^{-5}$ GeV$^{-2}$) is the Fermi 
coupling constant and $\cos\theta_{C}~(=0.9725)$ is the matrix 
element in Cabibbo-Kobayashi-Maskawa (CKM) matrix. The kinematic factors 
$u$ and $v$ are given by : $u,v=\Big(E_{\nu}+E_{\mu}~\pm~|{\bf q}|\Big)/(2~E_{\nu})$.
The pion decay constant is $f_{\pi}~(=0.93 ~m_{\pi})$ and 
${d\sigma(\pi^{+}A \rightarrow \pi^{+}A)}/{dt}$ is the pion-nucleus 
differential elastic cross section. The high energy approximation 
to the true minimal $Q^{2}$ is given by $Q^{2}_{m}=m^{2}_{\mu}~\nu/(E_{\nu}-\nu)$.
 The axial vector form factor can be defined 
as $G_{A}=m^{2}_{A}/(Q^{2}+m^{2}_{A})$  \cite{Berger:2008xs} with the axial vector
 meson mass $m_{A}$ (=0.95 GeV). The first term inside the brackets of 
Eq.~(\ref{marage1993}) corresponds to outgoing muons with 
negative helicity $($helicity nonflip$)$ whereas the second term is 
the helicity flip contribution which vanishes at $0^{\circ}$ scattering 
angle. 
The $\nu$ integration should be done in the range \cite{Paschos:2005km}
\begin{equation}
\rm max \Big(\xi \sqrt{Q^{2}} , \nu_{min} \Big) <   \nu   <  \nu_{max}~.
\end{equation}
  Here $\nu_{\rm min}$ and $\nu_{\rm max}$ are given in appendix. In our 
calculation, we use $\xi$ = 1 and 2.

\section{\bf Elastic pion nucleus cross section}
\subsection{\bf Berger-Sehgal (BS) Model}
In the Berger-Sehgal (BS) model~\cite{Berger:2008xs}, the elastic 
pion-nucleus scattering cross section is obtained from the pion-nucleon 
scattering cross section as 
\begin{equation}
\frac{d\sigma_{el}(\pi+A \rightarrow \pi+A)}{dt}=A^{2} \, \frac{d\sigma_{el}}
{dt}\Bigg|_{t=0} \, e^{-b t} F_{abs}~.
\label{pioncarbonxsection2014}
\end{equation}
The differential elastic pion-nucleon cross section in the forward direction 
is determined by the optical theorem as
\begin{equation}
\frac{d\sigma_{el}}{dt}\Bigg|_{t=0}=\frac{1}{16\pi} \, \Bigg(\frac
{\sigma^{\pi^{+} p}_{tot} + \sigma^{\pi^{-} p}_{tot}}{2}\Bigg)^{2}~,
\end{equation}
and the slope of the exponential $t$-distribution is calculated as
$ b=(1/3)R^{2}_{0}A^{2/3} $ with $R_{0}=1.057$ fm. $A$ is the mass number 
of the nucleus. $F_{abs}$ describes the average attenuation of pions in 
a nucleus,
\begin{equation}
F_{abs}=\exp\Bigg(-\frac{9 A^{\frac{1}{3}}}{16 \pi R^{2}_{0}} \, \sigma_{inel} \Bigg),
\end{equation}
where
\begin{equation}
\sigma_{inel}=\frac{\sigma^{\pi^{+} p}_{inel} + \sigma^{\pi^{-} p}_{inel}}{2} \, , \quad
\sigma_{inel} = \sigma_{tot} - \sigma_{el}~.
\end{equation}
The total elastic pion-nucleus cross section is calculated as :
\begin{equation}
\sigma_{el}(\pi+A \rightarrow \pi+A)=\frac{{A}^{2} F_{abs}}{16 \pi b} \,
\Bigg(\frac{\sigma^{\pi^{+} p}_{tot} + \sigma^{\pi^{-} p}_{tot}}{2} \Bigg)^{2}~.
\label{derivedpionnucleus}
\end{equation}
The pion-nucleon cross sections are taken from Particle Data Group
\cite{pdg2014}. We fit the pion-proton data with the superposition of 
a Breit-Wigner (BW) function and a Regge inspired term 
$a_{0}+a_{1}/\sqrt{p_{\pi}} + a_{2}\sqrt{p_{\pi}}$
up to 4 GeV. Above 4 GeV, the form $b_{0}+b_{1}/\sqrt{p_{\pi}} + b_{2}/{p_{\pi}}$ is taken.

\subsection{\bf Glauber model : Present}

 To calculate the differential elastic scattering cross section, the BS approach assumes
exponential $t$ dependence and models pion absorption with the term $F_{abs}$ in 
Eq.~(\ref{pioncarbonxsection2014}). In the Glauber model such assumptions are not required.
The scattering matrix $S_{l}$ is given by \cite{Shukla:2001mb}
\begin{equation}
S_{l} = \exp(i\chi(b)), \quad b k = \Big(l + \frac{1}{2}\Big),~
\end{equation}
where the Glauber phase shift $\chi(b)$ can be written as 
\begin{equation}
\chi(b) = \frac{1}{2}\sigma_{\pi N} \Big(\alpha_{\pi N} + i\Big) 
A T(b) \equiv \chi_{1} + i \, \chi_{2} ~.
\end{equation}
Here $\sigma_{\pi N}$ is the average total pion-nucleon cross section and 
$\alpha_{\pi N}$ is the ratio of real to imaginary part of the $\pi N$ scattering 
amplitude. 
In momentum space, $T(b)$ is defined as \cite{Shukla:2001mb}
\begin{equation}
T(b) = \frac{1}{2 \pi} \, \int J_{0}(q b) S(q) f_{\pi N}(q) q dq.
\end{equation}
Here $S(q)$ is the Fourier transform of the nuclear density $\rho(r)$
and $J_{0}(q b) = (1/2\pi)$ $\int \exp(-q b \cos\phi)d\phi$ is the cylindrical 
Bessel function of zeroth order. The function $f_{\pi N}(q)$ is the Fourier 
transform of the profile function $T(b)$ for $\pi N$ scattering which is taken as 
the Gaussian function of width $r_{0}$ as \cite{Shukla:2001mb}
\begin{equation}
T(b) = \frac{\exp\Big(-b^{2}/(2r^{2}_{0})\Big)}{2\pi r^{2}_{0}}~.
\end{equation}
Thus 
\begin{equation}
f_{\pi N} (q) = \exp\Big(\frac{-r^{2}_{0} q^{2}}{2}\Big).
\end{equation}
Here $r_{0}$ (= 0.6 fm) is the range parameter and may have a weak dependence on energy.
   
The pion-nucleus differential elastic cross section is calculated as 
\begin{equation}
\frac{d\sigma_{el}}{dt} = \frac{\pi}{k^{2}}|f(t)|^{2}, 
\label{dsigmadtpresent}
\end{equation}
where $f(t)$ is given by

\begin{eqnarray}
  f(t) = \frac{1}{2 i k} \sum \Big(2 l + 1\Big) \Big(S_{l}-1\Big) 
P_{l}\Big(\cos\theta\Big) .
\label{ftdsigmaeldtterm}
 \end{eqnarray}
Here $t = 4k^{2} \, \sin^{2}\theta/2$ and $k$ is the momentum of pion.
The total elastic cross section is calculated as 
\begin{equation}
\sigma_{el} = \frac{\pi}{k^{2}} \sum\limits_{l=0}^{\infty} \Big(2 l + 1\Big) 
\Big(S_{l}-1\Big)^{2}.
\label{sigmapresentelastic}
\end{equation}
Here 
$\Big(S_{l}-1\Big)^{2} = 1 + e^{-2\chi_{2}} - 2e^{-\chi_{2}}\cos\chi_{1}$. \\
The reaction cross section is calculated as \cite{Shukla:2001mb}
\begin{equation}
\sigma_{R} = \frac{\pi}{k^{2}} \sum\limits_{l=0}^{\infty} \Big(2 l + 1\Big) 
\Big(1 - |S_{l}|^{2}\Big)~.
\end{equation}
  The total cross section is calculated as $\sigma_{tot} = \sigma_{el} + \sigma_{R}$.
The values of $\alpha_{\pi N}$ for pion-carbon scattering are extracted by fitting the 
total pion-carbon cross section (obtained by experiment at low pion energies) and assumed 
to be $A \, \sigma_{\pi-N}$ at higher pion energies. 
 For carbon, we obtain $\alpha_{\pi N}$= 1.5 for $E_\pi < 5 $ GeV and 1.4 for $E_\pi > 5 $ GeV.
For iron, we obtain $\alpha_{\pi N}$= 1.8 for $E_\pi < 10 $ GeV and 1.75 for $E_\pi > 10 $
GeV.

 The nuclear density function for carbon nuclei is taken as the harmonic 
oscillator type as given by 
\begin{eqnarray}
\rho(r) &=& \rho_{0} \Big(1 + \alpha \, \frac{r^{2}}{a^{2}}\Big) 
\exp\Big(-\frac{r^{2}}{a^{2}}\Big), \quad
\rho_{0} = \frac{1 + 1.5 \alpha}{(\sqrt{\pi}\,a)^{3}}~.
\label{carbonnucleardensity}
\end{eqnarray}
 The values of $\alpha$~(= 1.247 fm) and $a$~(= 1.649 fm) are taken from the 
Ref.~\cite{DeJager:1974liz}.

 The nuclear density function for iron nuclei is taken as a 
two-parameter Fermi form as
\begin{eqnarray}
\rho(r) &=& \frac{\rho_{0}}{1 + \exp\Big(\frac{r-c}{d}\Big)}, \quad
\rho_{0} = \frac{3}{4 \pi c^{3} \, \Big(1 + \frac{\pi^{2} d^{2}}{c^{2}}\Big)}~.
\label{ironnucleardensity}
\end{eqnarray}
The values of $c$~(=4.111 fm) and $d$~(=0.558 fm) are taken from 
the Ref.~\cite{DeJager:1987qc}. 

  The BS approach assumes that the differential pion-nucleus cross section is an 
exponential distribution in $t$. 
  The factor $F_{abs}$ in Eq.~(\ref{pioncarbonxsection2014}) which represents the 
absorption of pions in a nucleus has been obtained in a geometrical model 
assuming the nucleus as a sharp sphere. In the Glauber (present) model, the 
$t$ distribution given by Eq.~(\ref{dsigmadtpresent}) with Eq.~(\ref{ftdsigmaeldtterm})  
is obtained from scattering theory. Realistic forms of nuclear densities are used as 
given by Eqs.~(\ref{carbonnucleardensity}) and (\ref{ironnucleardensity}).

\section{\bf Results and discussions}

\begin{figure}
\includegraphics[width=0.98\linewidth]{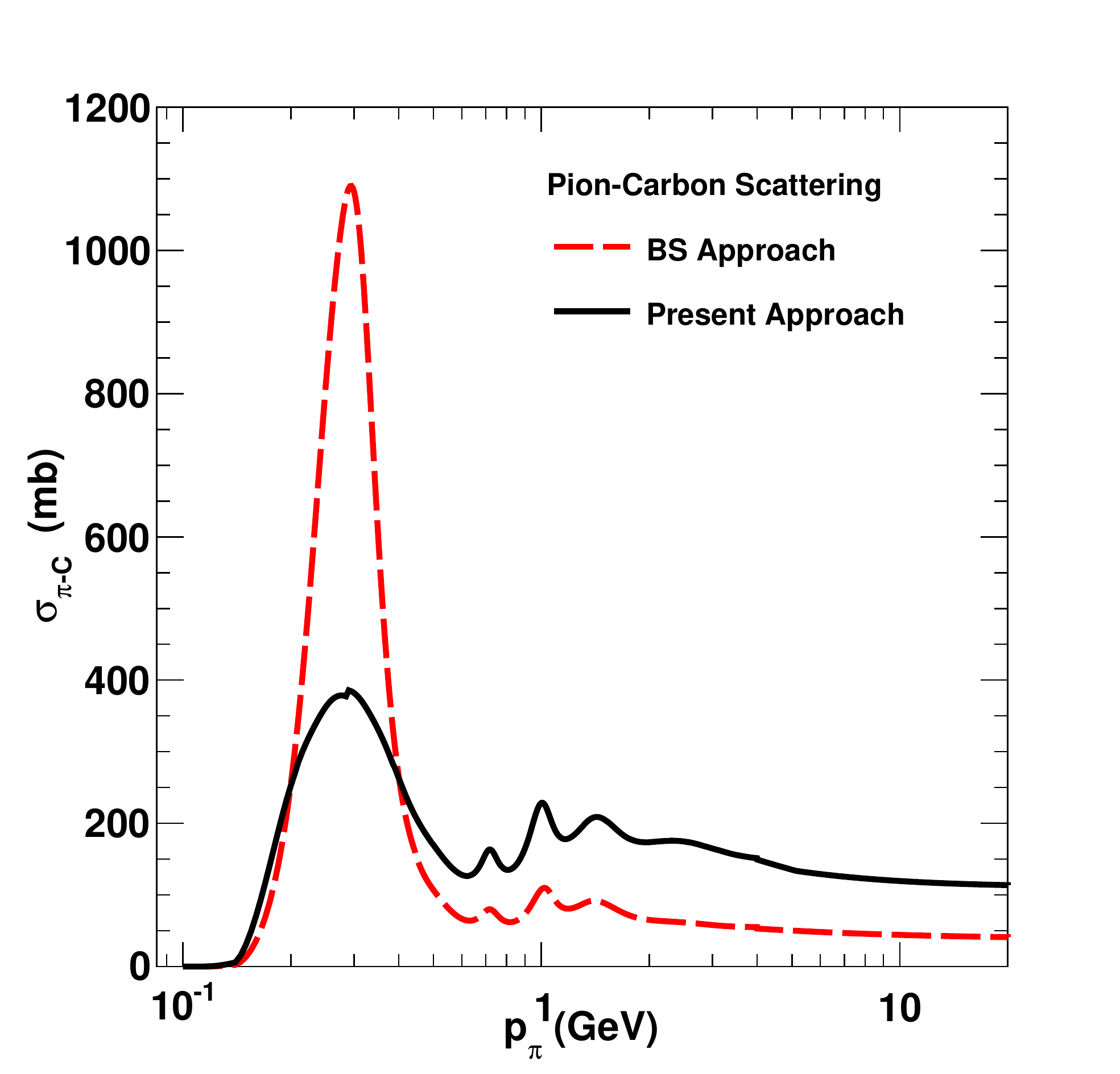}
\caption{(Color online) Total elastic cross section $\sigma_{\pi C}$ of pion-carbon 
scattering as a function of pion momentum $p_{\pi}$.}
\label{figure2pioncarbonderived}
\end{figure}

\begin{figure*}
\includegraphics[width=0.48\linewidth]{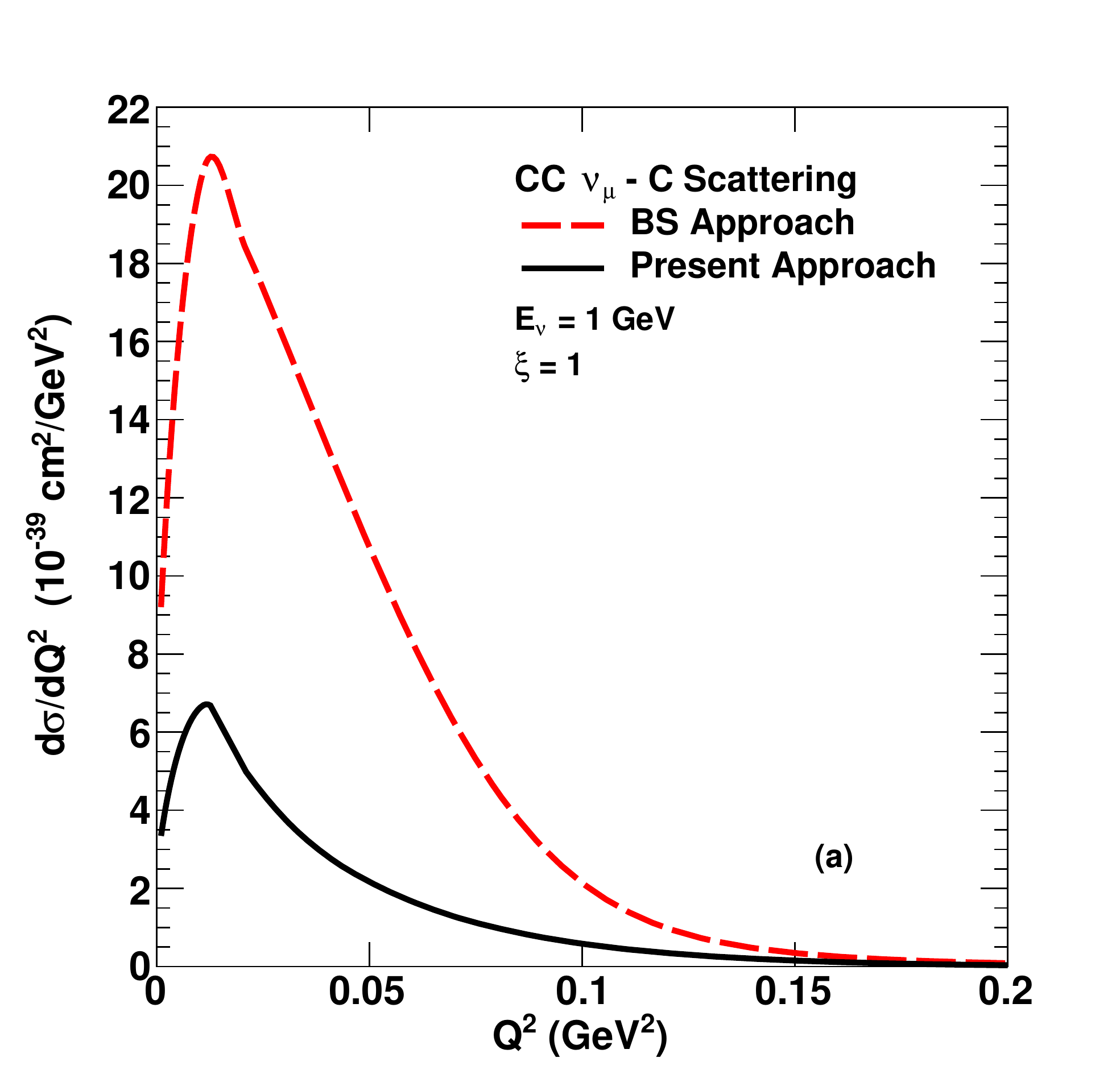}
\includegraphics[width=0.48\linewidth]{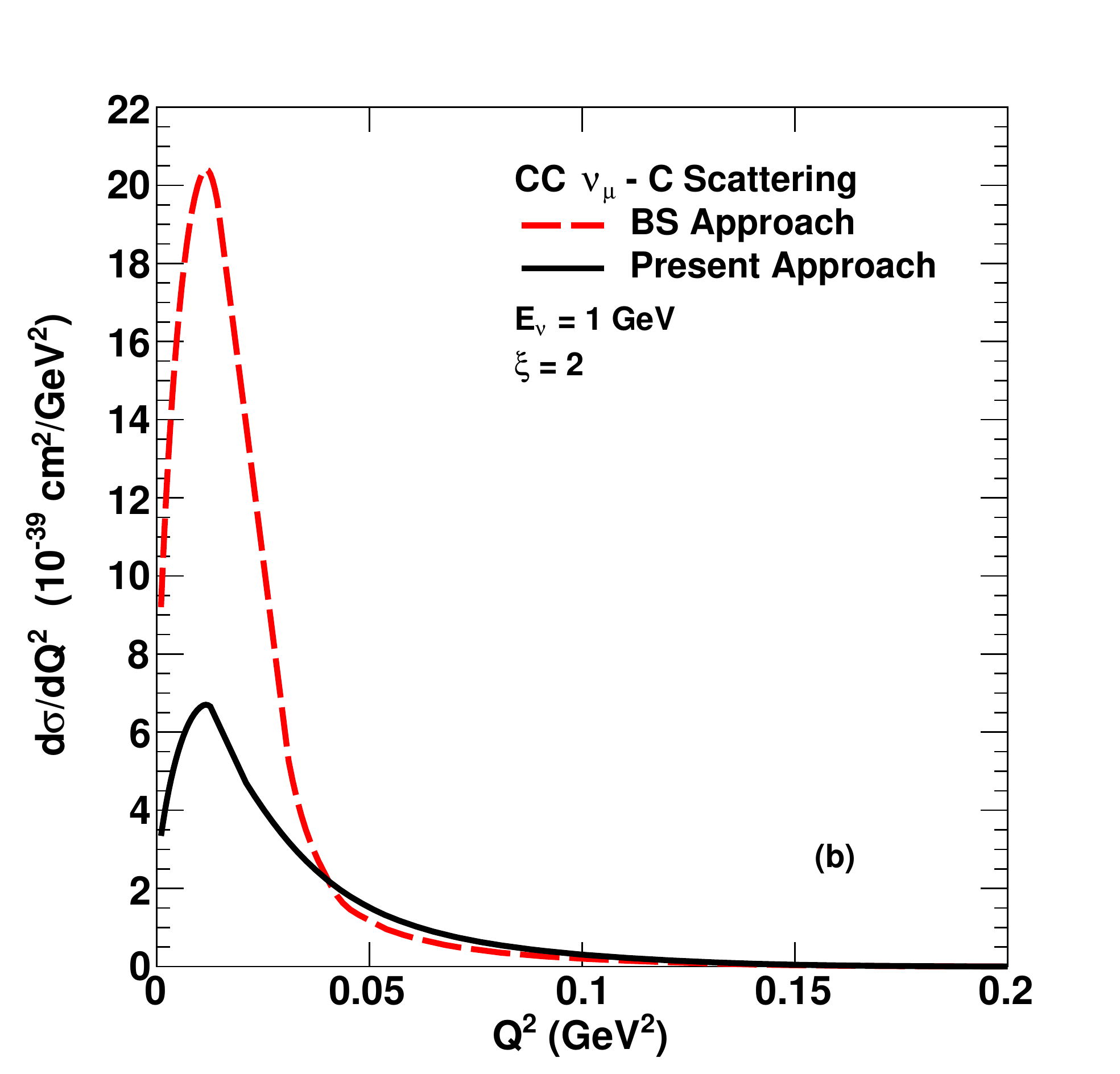}
\caption{(Color online) Differential cross section $d\sigma/dQ^{2}$ for the charged 
current coherent $\nu_{\mu}-C$ scattering as a function of $Q^{2}$ obtained using the 
PCAC-based model (BS and Present) at neutrino energy 1 GeV for (a) $\xi$=1 and (b) $\xi$=2.} 
\label{figure3kmneutrinoonezaione}
\end{figure*}

\begin{figure*}
\includegraphics[width=0.48\linewidth]{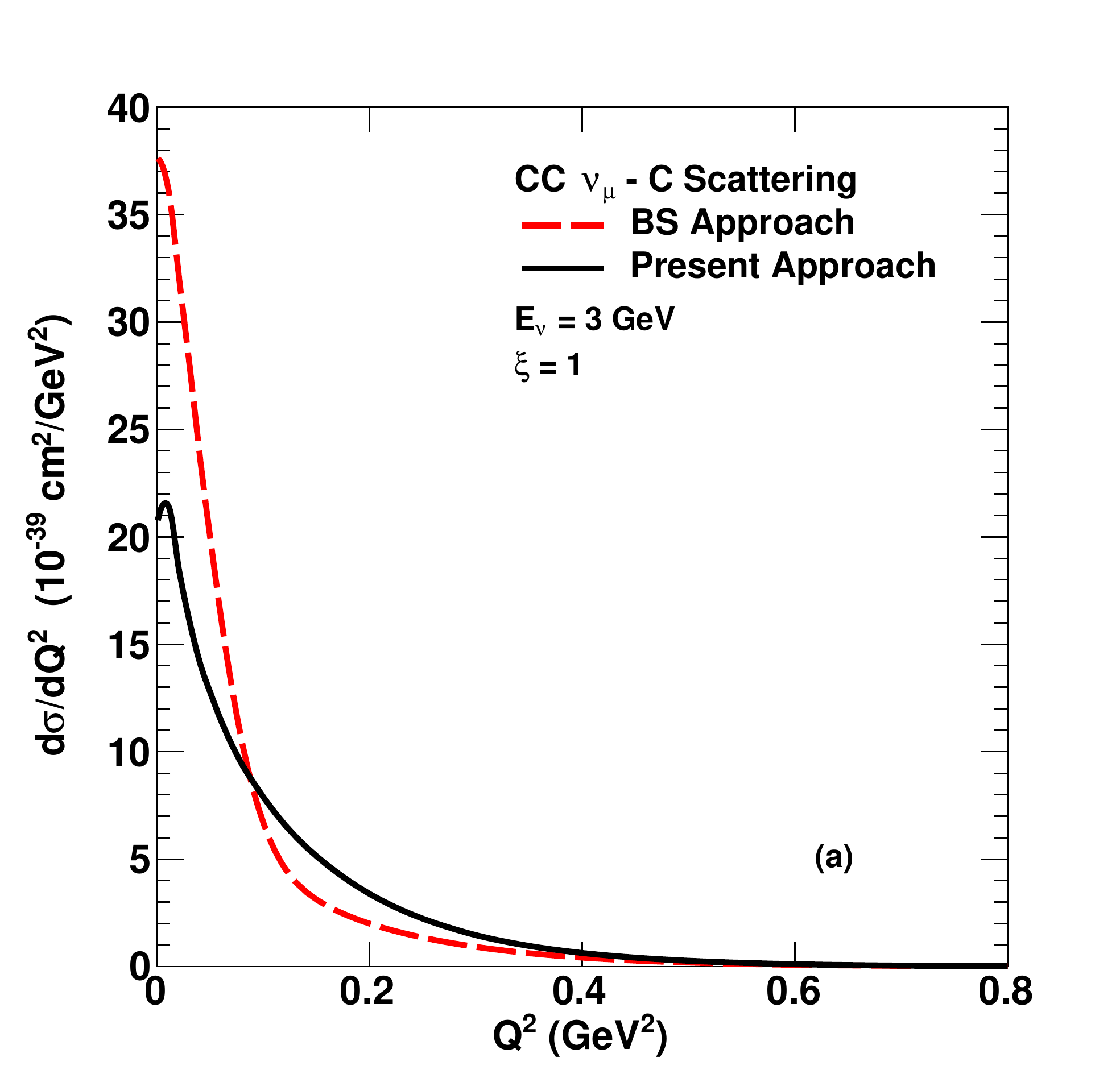}
\includegraphics[width=0.48\linewidth]{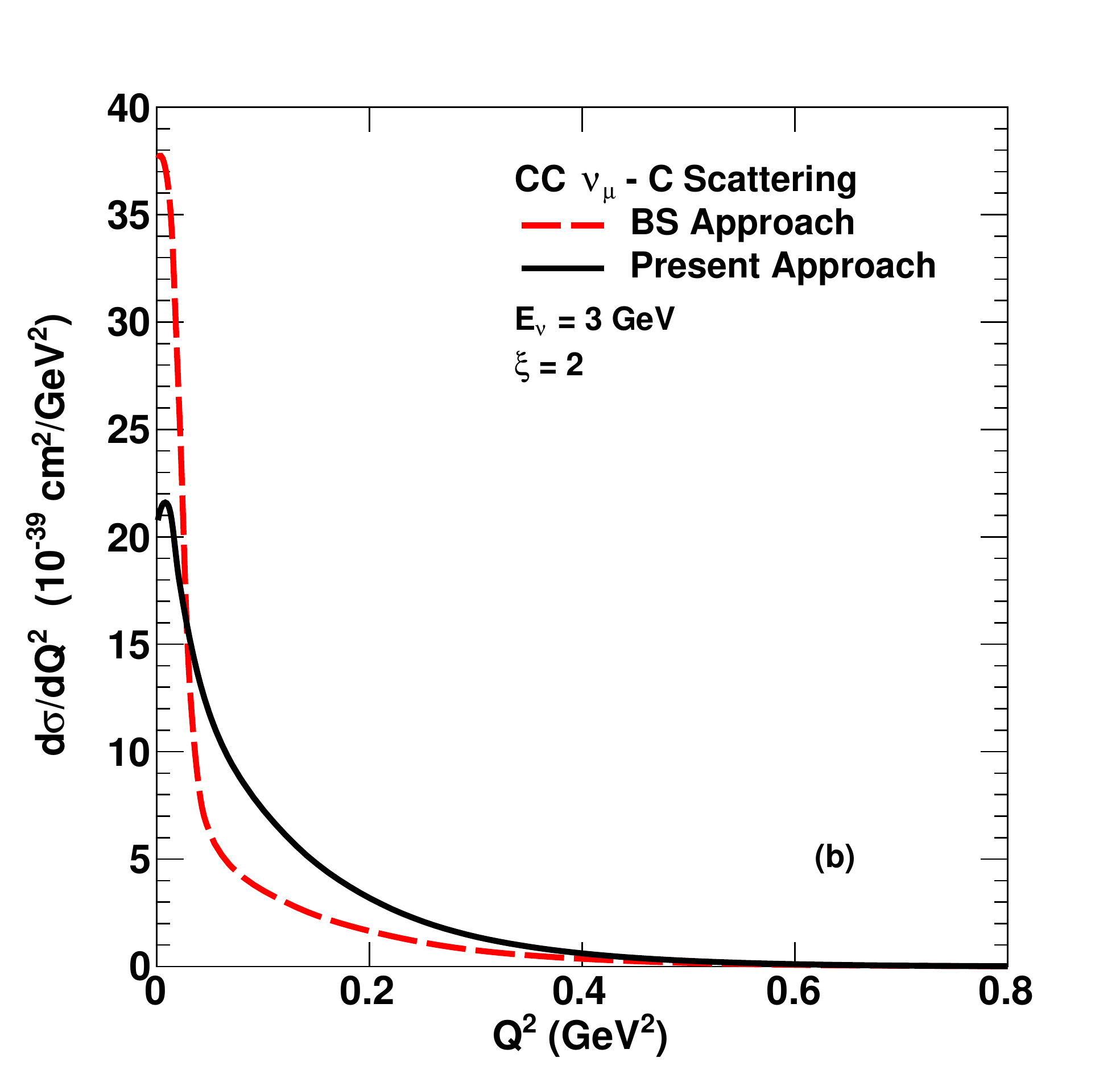}
\caption{(Color online) Differential cross section $d\sigma/dQ^{2}$ for the 
charged current coherent $\nu_{\mu}-C$ scattering as a function of $Q^{2}$ 
obtained using the PCAC-based model (BS and Present) 
at neutrino energy 3 GeV for (a) $\xi$=1 and (b) $\xi$=2. } 
\label{figure4kmneutrinothreezaione}
\end{figure*}

  Figure~\ref{figure2pioncarbonderived} shows total pion Carbon elastic scattering 
cross section $\sigma_{\pi C}$ as a function of pion momentum $p_{\pi}$. 
The broken line is obtained using the Berger-Sehgal approach  according to 
Eq.~(\ref{derivedpionnucleus}) while the solid line represents the Glauber calculation 
according to Eq.~(\ref{sigmapresentelastic}). The $\sigma_{\pi C}$ according to the BS approach
sharply peaks at $p_{\pi}$ = 0.3 GeV. Above 1 GeV pion momentum, $\sigma_{\pi C}$ 
from both the approaches vary slowly as a function of pion momentum.

 We calculate differential cross section $d\sigma/dQ^{2}$ for the charged current 
coherent neutrino-carbon scattering using the PCAC-based model at different energies 
to study the effect of parameter $\xi$ for both the approaches used in the present work. 
  Figure~\ref{figure3kmneutrinoonezaione}(a) shows $d\sigma/dQ^{2}$ as a function of 
the square of momentum transfer $Q^{2}$ obtained using the PCAC-based model (BS and Present) 
at 1 GeV neutrino energy for $\xi$=1. The cross sections from both the approaches peak 
at low $Q^{2}$ with the present calculation giving smaller cross sections than the 
BS calculations.
Figure~\ref{figure3kmneutrinoonezaione}(b) shows $d\sigma/dQ^{2}$ 
as a function of $Q^{2}$ obtained using the PCAC-based model 
at 1 GeV neutrino energy but for $\xi$=2. 
 Here also the present calculation gives smaller cross sections than 
the BS calculation while at higher values of $Q^{2}$ it crosses the BS calculations.
 From Figs.~\ref{figure3kmneutrinoonezaione}(a) and \ref{figure3kmneutrinoonezaione}(b)
we can see that with increasing the value of $\xi$, the cross section is reduced in 
both approaches.

\begin{figure*}
\includegraphics[width=0.48\linewidth]{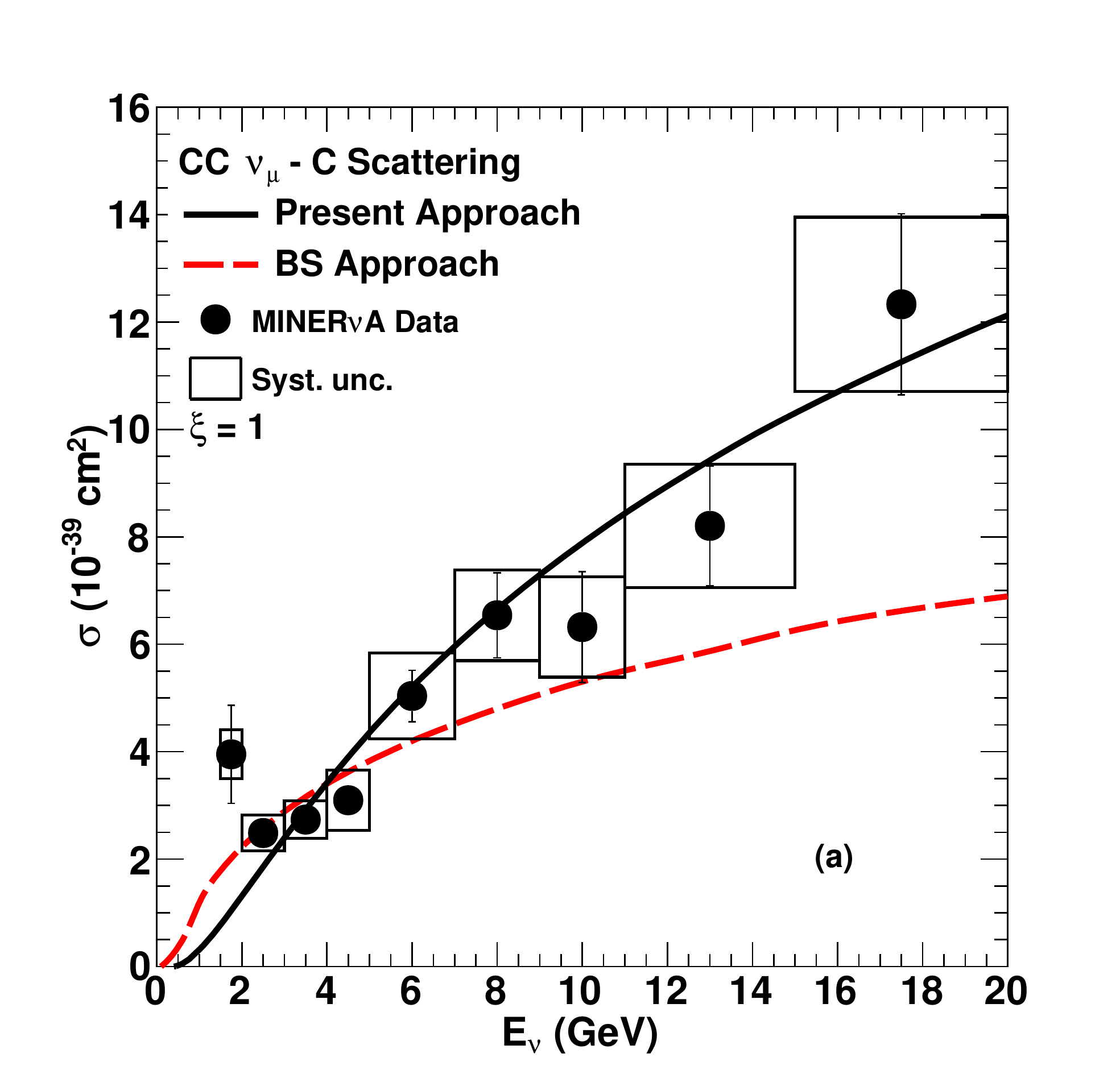}
\includegraphics[width=0.48\linewidth]{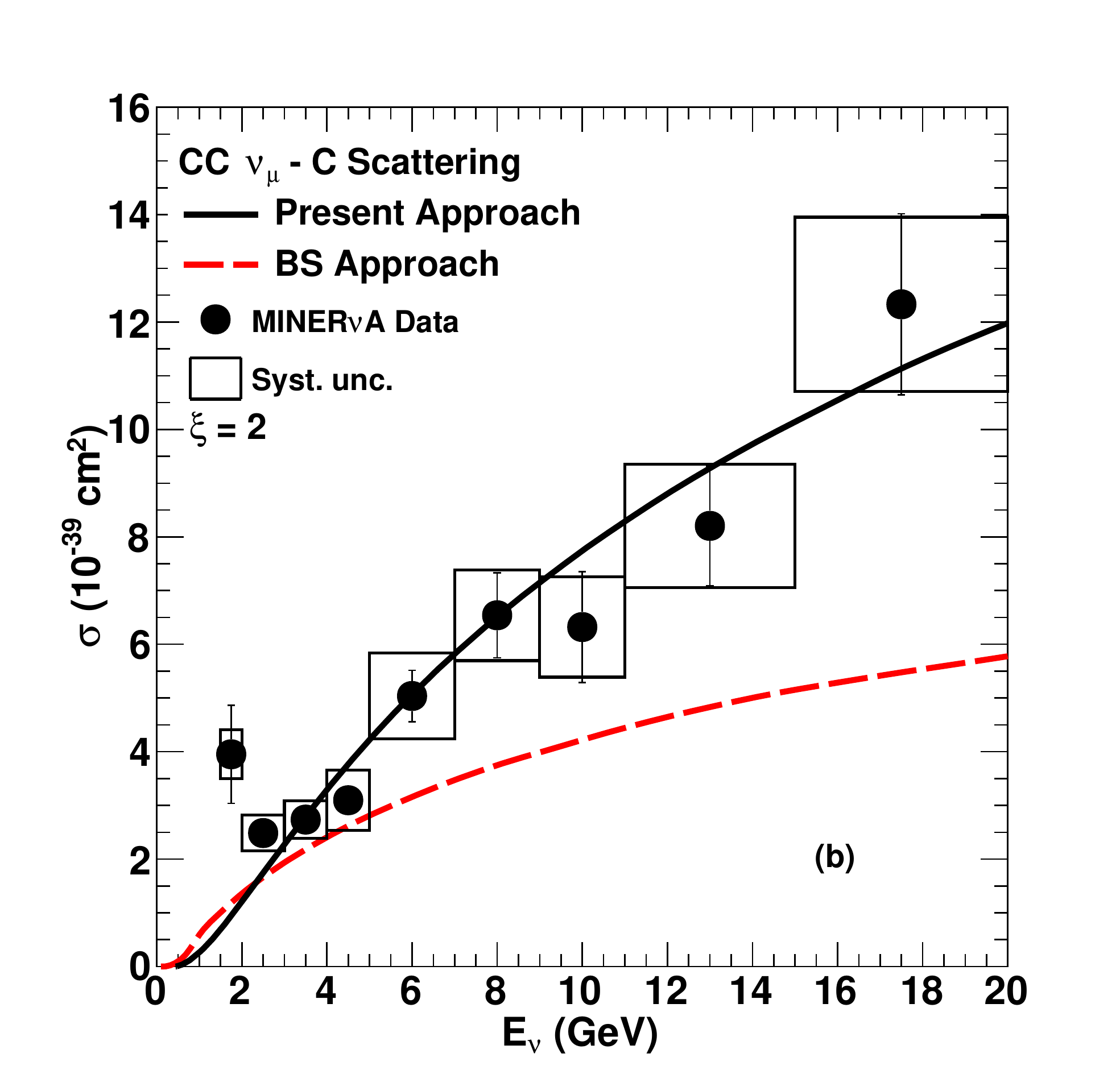}
\caption{(Color online) Total cross section $\sigma$ for the charged current coherent $\nu_{\mu}-C$ 
scattering as a function of neutrino energy $E_{\nu}$ obtained using the PCAC-based 
model (BS and Present) for (a) $\xi$=1 and (b) $\xi$=2. } 
\label{figure5kmxsectionzaione}
\end{figure*}

\begin{figure*}
\includegraphics[width=0.48\linewidth]{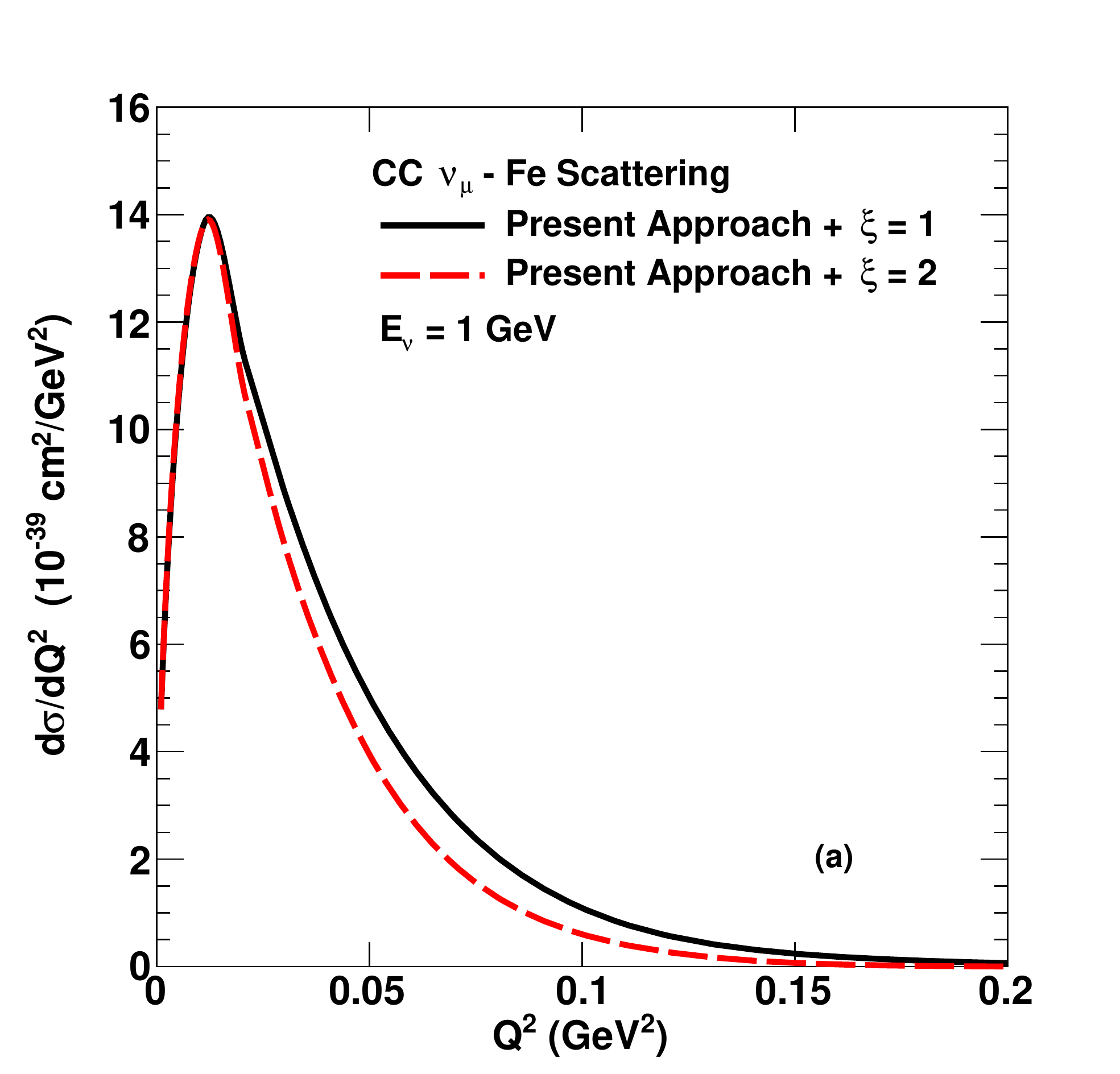}
\includegraphics[width=0.48\linewidth]{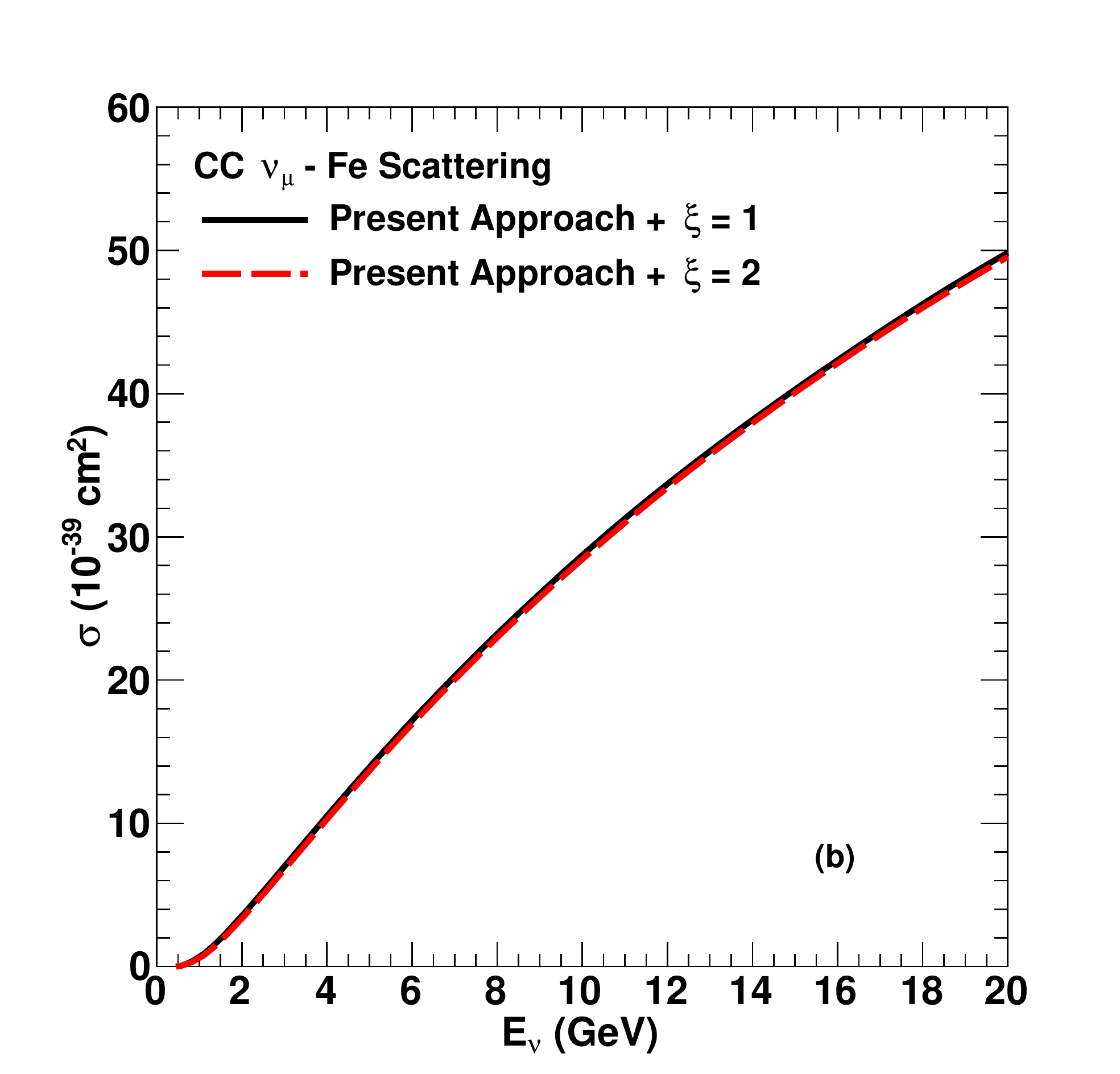}
\caption{(Color online) (a) Differential cross section $d\sigma/dQ^{2}$ for the charged current 
coherent $\nu_{\mu}-Fe$ scattering as a function of  $Q^{2}$ obtained using the PCAC-based 
model (Present) for $\xi$=1 and 2 at $E_\nu=1$ GeV.
(b) Total cross section $\sigma$ for the charged current coherent $\nu_{\mu}-Fe$ 
scattering as a function of neutrino energy $E_{\nu}$ obtained using the PCAC-based 
model (Present) for $\xi$=1 and 2.} 
\label{figure6ironneutrinoonezaionetwo}
\end{figure*}

Figure~\ref{figure4kmneutrinothreezaione}(a) shows $d\sigma/dQ^{2}$ as a 
function of $Q^{2}$ at 3 GeV neutrino energy for $\xi$=1. 
 At low $Q^{2}$, the present calculation gives smaller cross sections than the BS 
calculation while at higher values of $Q^{2}$ the present calculation gives larger
cross section. 
   Figure~\ref{figure4kmneutrinothreezaione}(b) shows $d\sigma/dQ^{2}$ as a 
function of $Q^{2}$ at  3 GeV neutrino energy for $\xi$=2. For higher values 
of $\xi$, both calculations sharply peak at low  $Q^{2}$ with the sharpness of 
the peak greater for the BS approach.

Figure~\ref{figure5kmxsectionzaione}(a) shows total cross sections $\sigma$ for the 
charged current coherent neutrino-carbon scattering as a function of neutrino 
energy $E_{\nu}$ obtained using the PCAC model (BS and present) for  $\xi$=1. The 
calculations are compared with the data recorded by the MINER$\nu$A experiment 
\cite{Higuera:2014azj}. At lower neutrino energies ($E_{\nu}~\leq$ 4 GeV), both calculations 
are compatible with the data. At higher energies ($E_{\nu}~\geq$ 5 GeV), the present 
approach gives better description of the data.
  Figure~\ref{figure5kmxsectionzaione}(b) shows the total cross section $\sigma$ for the 
charged current coherent neutrino carbon scattering as a function of neutrino 
energy $E_{\nu}$ obtained using the PCAC model (BS and present) for  $\xi$=2. The 
calculations are compared with the data recorded by the MINER$\nu$A experiment 
\cite{Higuera:2014azj}. At lower energies ($E_{\nu}~\leq$ 4 GeV), both calculations 
are compatible with the data within experimental error. With increased value of $\xi$, 
the present approach gives excellent description of the data in all energy range.

  We also give predictions for cross sections in neutrino-iron coherent scattering. 
Figure~\ref{figure6ironneutrinoonezaionetwo}(a) shows differential cross sections 
$d\sigma/dQ^{2}$ for the charged current neutrino-iron coherent scattering as a function of 
$Q^{2}$ obtained using the PCAC-based model (present) at 1 GeV neutrino energy for $\xi$=1 and 2. 
For the case of iron we do not find drastic change with changing the value of $\xi$.
  Figure~\ref{figure6ironneutrinoonezaionetwo}(b) shows total cross section $\sigma$ for the 
charged current coherent neutrino-iron scattering as a function of neutrino energy 
$E_{\nu}$ obtained using the PCAC-based model (present) for $\xi$=1 and 2. 
With increasing value of $\xi$, the total cross section is slightly reduced and
is not very sensitive to the value of $\xi$.

\section{Conclusion}

 We presented the differential and integrated cross sections of coherent pion production 
in neutrino-nuclei scattering 
using the formalism based on partially conserved axial current (PCAC) theorem 
which relates the neutrino-nucleus cross section to the pion-nucleus elastic cross section.
We study the behavior of the cross sections as a function of neutrino energy and the 
parameters of the model.
  The pion-nucleus elastic cross section is calculated using the Glauber model in terms of
the pion-nucleon cross sections obtained by parameterizing the experimental data. 
The results obtained using this approach have been compared with those obtained using 
the Berger-Sehgal approach.
 The calculated integrated cross sections are compared with the measured data.
 At lower energies ($E_{\nu}~\leq$ 4 GeV) both the approaches are compatible with the data. 
The present approach gives good description of the data in all energy range.
 Predictions for differential and integrated cross sections for the coherent pion 
productions in neutrino-iron scattering using the above formalism are also made.

\section{\bf Appendix : Kinematic Limits For Integrations}
The integration range over $t$ is given as \cite{Paschos:2005km}:
\begin{equation}
|t_{min}|~<~ -t ~< ~ 0.05~GeV^{2},
\label{trange005}
\end{equation}
where
\begin{eqnarray}
t_{min} &=& \frac{(Q^{2}+m^{2}_{\pi})^{2} \, - \, \Big[\sqrt{\lambda_{1}}
\, - \, \sqrt{\lambda_{2}}\Big]^{2}}{4W^{2}} ,\nonumber \\
 & \approx & - \, \Bigg(\frac{Q^{2}+m^{2}_{\pi}}{2\nu}\Bigg)^{2}.~
\label{tminlimit}
\end{eqnarray}
Here
\begin{eqnarray}
\lambda_{1} &=& \lambda(W^{2},-Q^{2},M^{2}_{A})~, \nonumber \\
\lambda_{2} &=& \lambda(W^{2},m^{2}_{\pi},M^{2}_{A})~,\nonumber \\
\lambda(a,b,c) &=& a^{2}+b^{2}+c^{2}-2 a b-2 a c-2 b c~.
\end{eqnarray}

The kinematical minimum and maximum values of $\nu$ are :
\begin{equation}
\nu_{min}=\frac{(W^{2}_{min}\, +\, Q^{2}\, -\, M^{2}_{A})}{2M_{A}},
\end{equation}
\begin{equation}
\nu_{max}=\frac{(W^{2}_{max}\, +\, Q^{2}\, -\, M^{2}_{A})}{2M_{A}},
\end{equation}
where
\begin{eqnarray}
 W^{2}_{min}= (M_{A}+m_{\pi})^{2},   
\end{eqnarray}
\begin{eqnarray}
W^{2}_{max} &=& \Bigg\{\frac{1}{4}\, s^{2}\, \Bigg(1\, -\, \frac{M^{2}_{A}}{s}\Bigg)^{2}\, 
\Bigg(1\, -\, \frac{m^{2}_{\mu}}{s}\Bigg)   \nonumber \\
&-& \Bigg[Q^{2}\, -\, \frac{s}{2}\, \Bigg(1\, -\, 
\frac{M^{2}_{A}}{s}\Bigg)\, +\, \frac{m^{2}_{\mu}}{2}\, \Bigg(1\, +\, \frac{M^{2}_{A}}{s} 
\Bigg)\Bigg]^{2}\Bigg\} \nonumber \\
\, & \times &\,  \Bigg(1\, -\, \frac{M^{2}_{A}}{s}\Bigg)^{-1}\, (Q^{2}\, +\, m^{2}_{\mu})^{-1}.
\end{eqnarray}
Here  $s = M^{2}_{A}\, +\, 2M_{A}\, E_{\nu}$.\\
The $\nu$ integration should be done in the range \, \cite{Paschos:2005km}
\begin{equation}
\rm max \Big(\xi\, \sqrt{Q^{2}}\, ,\,  \nu_{min} \Big)\,  < \,  \nu \,  < \, 
\nu_{max}.
\end{equation}

The kinematically allowed minimum value of $Q^{2}$  is :
\begin{eqnarray}
Q^{2}_{min}&=&\frac{(s-M^{2}_{N})}{2}\, \Bigg[1\, -\, \lambda^{\frac{1}{2}}\Bigg
(1,\frac{m^{2}_{\mu}}{s},\frac{W^{2}_{min}}{s}\Bigg)\Bigg]\, \\
&-&\frac{1}{2}\, \Bigg[
W^{2}_{min}\, +\, m^{2}_{\mu} - \frac{M^{2}_{A}}{s} (W^{2}_{min}\, -\, m^{2}_{\mu})\Bigg]
\end{eqnarray}
$Q^{2}$ region for coherent scattering is :
\begin{equation}
Q^{2}_{min} \, <\, Q^{2}\, \lesssim\, 2.0\, GeV^{2}.
\end{equation}


\end{document}